\documentclass[12pt]{article}
\usepackage{epsfig}

\parskip=\medskipamount 
plus.3em minus.5em \abovedisplayshortskip=.5em plus.2em minus.4em
\renewcommand{\thesection}{\arabic{section}}

\catcode`@=11

\@addtoreset{equation}{section} \@addtoreset{equation}{subsection}
\def\theequation{\ifnum\value{section}=0 \arabic{equation}\ignorespaces
\else \ifnum\value{section}=-1 A.\arabic{equation}\ignorespaces
\else \ifnum\value{subsection}=0
\thesection.\arabic{equation}\ignorespaces \else
\thesection.\arabic{subsection}.\arabic{equation}\ignorespaces
                             \fi
                        \fi
                   \fi}

{\catcode`\'=\active \def'{{}^\bgroup\prim@s}}

\catcode`@=12

\newcommand{\bq}{\begin{equation}}
\newcommand{\be}{\begin{equation}}
\newcommand{\fq}{\end{equation}}
\newcommand{\ee}{\end{equation}}
\newcommand{\bqr}{\begin{eqnarray}}
\newcommand{\beqs}{\begin{eqnarray}}
\newcommand{\fqr}{\end{eqnarray}}
\newcommand{\eeqs}{\end{eqnarray}}

\newcommand{\rf}[1]{(\ref{#1})}

\def\bop#1{\setbox0=\hbox{$#1M$}\mkern1.5mu
    \vbox{\hrule height0pt depth.04\ht0
    \hbox{\vrule width.04\ht0 height.9\ht0 \kern.9\ht0
    \vrule width.04\ht0}\hrule height.04\ht0}\mkern1.5mu}

\begin{document}
\thispagestyle{empty}

\begin{flushright}
\begin{tabular}{l}
hep-th/0508221 \\
\end{tabular}
\end{flushright}

\vskip .6in
\begin{center}

{\bf Mass Patterns in the Fermion Spectrum}

\vskip .6in

{\bf Gordon Chalmers}
\\[5mm]

{e-mail: gordon@quartz.shango.com}

\vskip .5in minus .2in

{\bf Abstract}

\end{center}

All of the known fermions fit a simple formula which is presented here.  
The simple formula appears to indicate a topological origin to mass 
generation, and it is globally accurate to an approximate percent.  
Instantons or a quantized two Higgs mechanism, as in the MSSM model, 
could give a simple origin to the masses.

\vfill\break

\vskip .2in

The origin of the fermion masses has been investigated for many years.  
Several 
mass relations including those of the form $m_1\sim m_2^2$ have been proposed, 
but without dynamical meaning.  Remarkably, there is a simple formula that 
matches with the known spectrum, in the scales from an MeV to almost a 
TeV, 

\bqr 
m_f = 10^{n-4} ~{\rm GeV} \pm 2^i 5^j ~{\rm MeV} \ . 
\label{massformula} 
\fqr 
The formula in \rf{massformula} matches all of the fermions, over the scales 
from an MeV to a TeV to an approximate one percent.  

Furthermore, the mass formula appears to fit a few data points which are 
published in the literature.  

The masses of the known fermions in the standard model are represented in 
the following mass table, 

\bqr 
\pmatrix{ & {\rm GeV}  &   {\rm rem}      &   10^n   &  .002 ~{\rm rem} \cr 
 {\rm up}    &   .004    & .003           & .001     &     1.5         \cr  
 {\rm down}  &   .008        & -.002          &  .01     &    -1           \cr  
 {\rm charm} &   1.5         & .5             &   1.     &     250         \cr   
 {\rm strange} &  .15        &  .05           &   .1     &     25          \cr     
 {\rm top}   &   176         & 76             &  100     &     25500*1.5   \cr 
 {\rm bottom} &  4.7         & -5.3           &   10     &     -2500       
}
\fqr 
and for the leptons, 
\bqr  
\pmatrix{ 
{\rm electron} & .005  & .01 &  -.5   &  250  \cr 	
{\rm muon}     &  .11  &  .1 &  .01   &  5   \cr 
{\rm tau}      & 1.7   &  1  &  .7    &  250*1.4     \ . 
}
\fqr 
The remainder is the mass after subtracting $10^n$, and their values 
are presented in units of $.002$ GeV; the latter indicates some kind 
of number similarities.  
The mass and the remainder from the numbers $10^n$, in GeV, are given.  Note 
that the remainders fit the pattern of $25*10^m$.   These number relations 
have apparently not appeared in the literature.  

The second terms, that is the remainders, can be fit by $2^i 5^j$ with 
$i$ and $j$ integers.  To a percent, the formula involving the $2$s and $5$s 
fit the known particle masses.  The masses are simply, 

\bqr 
10^{n-4} \pm 2^m 5^i ~10^{-3} ~{\rm GeV} \ , 
\fqr 
or in MeV, 
\bqr 
10^{n-1} +/- 2^m 5^i ~{\rm MeV}  \qquad  n=1,2,\ldots 6  \ .
\fqr 
The masses of the quarks are summarized in the table, 
\bqr  
\pmatrix{
{\rm quark} &  n  &  \pm & m  & i  &  - &   + \cr  
{\rm u}  &   1  &      &1 & 0   & -1 &  1 \cr 
{\rm d}  &   2  &  -   &1 & 0   & -1 &  1 \cr    
{\rm c}  &   4  &      &2 & 3   &  1 &  5 \cr 
{\rm s}  &   3  &      &1 & 2   &  1 &  3 \cr 
{\rm t}  &   6  &      &4 & 5   &  1 &  9 \cr
{\rm b}  &   5  &  -   &3 & 4   &  1 &  7 
}  \ . 
\fqr 
As mentioned, the masses are reproduced to an approximate percent, which 
appears to be almost as good as current experimental observations.  The 
$2^m 5^i$ contributions appear to scale with the $10^n$ number in a 
linear fashion.  

As a point of reference, there are only $27$ numbers out of $999$ of the form 
$2^m 5^i$.  (Recall that the chiral N=1 multiplet of ${\rm E}_8$ has $27$ 
entries, as a point of reference.)  

The robustness of the masses in accord with the mass formulae is clear.  The 
possible extension to broken supersymmetric theories is interesting, and could 
merit further investigation in experimental data.  Anomalous events might be 
present in the data that follow this pattern.  It is worth pointing out that 
some data could involve a factor of $3$, in addition to the $2^m 5^i$ 
factoring.  

The explanation of the mass formula, involving the two terms, does suggest 
either a two-Higgs contribution or instantonic ones (perhaps along the 
lines in a local sense as in \cite{Chalmers}).  The MSSM does have two 
Higgs modes, but the quantization of these modes is not obvious in 
perturbation theory.  Maybe the quantization already indicates the 
presence of extra dimensions through quantized Yukawa couplings.

\end{document}